\documentstyle[twoside,fleqn,espcrc2]{article}

\def\bc{\begin{center}}
\def\ec{\end{center}}
\def\be{\begin{equation}}
\def\ee{\end{equation}}
\def\myappendix{\par
 \setcounter{section}{0}
 \setcounter{subsection}{0}
 \setcounter{equation}{0}
 \setcounter{table}{0}
 \def\appendixname{Appendix}
 \def\appesection{\setcounter{equation}{0}\section}
 \def\@thesection{\Alph{section}}
 \def\thesection{\appendixname\hskip 1.10ex\Alph{section}}
 \def\thesubsection{\@thesection.\arabic{subsection}}
 \def\theequation{\@thesection.\arabic{equation}}
 \def\thetable{\@thesection.\arabic{table}}}

\newcommand{\labar}{\overline{\Lambda}}
\newcommand{\lb}{\overline{\Lambda}}
\newcommand{\lle}{\lambda_{1}}
\newcommand{\lc}{\lambda_{2}}
\newcommand{\fb}{f_{B}}
\newcommand{\bb}{B^{0}-\bar{B}^{0}}
\newcommand{\mb}{\overline{m}_{b}(\overline{m}_{b})}

\newcommand{\beq}{\begin{equation}}
\newcommand{\eeq}{\end{equation}}
\newcommand{\beqn}{\begin{eqnarray}}
\newcommand{\eeqn}{\end{eqnarray}}
\newcommand{\mfrac}[2]{\frac{\textstyle #1}{\textstyle #2}}

\def\vdir{v\kern-7.8pt\Big{/}}
\def\pdir{p\kern-7.8pt\Big{/}}

\title{B Physics on the Lattice:  $\lb$, $\lle$, $\mb$, $\lc$, $\bb$ mixing,
$\fb$ and all that.}

\author{V.~Gim\'enez\address{Dep. de Fisica Teorica and IFIC, Univ. de Valencia,\\
Dr. Moliner 50, E-46100, Burjassot, Valencia, Spain.}
\thanks{Talk presented by V.~Gim\'enez}
G.~Martinelli\address{I.N.F.N. and Dipartimento di Fisica dell' Universit\'a 
di Roma ``La Sapienza'' \\ Piazza Aldo Moro 2, I-00185 Roma, Italy} and
C.T.~Sachrajda\address{Theory Division, CERN, 1211 Geneva 23, 
Switzerland.}}
\begin{document}
\begin{abstract}
We present a short review of our most recent high statistics lattice
determinations in the HQET of the following important
parameters in B physics:
the B--meson binding energy, $\labar$ and the kinetic energy of the
b quark in the B meson, $\lle$, which due to the presence of power
divergences require a non--perturbative renormalization to be defined;
the $\overline{MS}$ running mass of the b quark, $\mb$;
the $B^{*}$--$B$ mass splitting, whose value in the HQET is determined
by the matrix element of the chromo--magnetic operator between B meson states,
$\lc$;
the B parameter of the $B^{0}$--$\bar{B}^{0}$ mixing, $B_{B}$, and
the decay constant of the B meson, $f_{B}$. All these quantities have been
computed using a sample of $600$ gauge field configurations on a
$24^{3}\times 40$ lattice at $\beta=6.0$. For $\labar$ and $\mb$, we obtain
our estimates by combining results from three independent lattice simulations
at $\beta=6.0$, $6.2$ and $6.4$ on the same volume.
\end{abstract}
\maketitle
\section{The calculation of $\lb$ and $\mb$.}
\label{labardef}

The presence of renormalon singularities in the pole mass, $m_b^{pole}$,
implies that the intuitive definition $\labar\equiv m_B - m_b^{pole}$ does
not correspond to a physical quantity.
Thus, if $\labar$ is to be used in phenomenological applications, a definition
free of renormalon ambiguities has to be given.
One can define the B--meson binding energy from the
experimentally measured value of some physical quantity, for example the
inclusive
$B \rightarrow X\, l\, \bar{\nu}_l$ lepton spectrum. It is important to
notice that such a definition, although correct, depends on the order of
the perturbation series that determines the theoretical prediction for the
experimental quantity.

Lattice simulations of the HQET provide the
opportunity to compute $\labar$ non--perturbatively by using the definition
of $\labar$ proposed in ref.\cite{gmdef} which is free of
renormalon ambiguities. 
On the lattice, renormalons are absent but matrix elements of
lattice operators may contain power ultraviolet divergences. Consider
the two--point correlation function  $C(t)$ for large values of $t$
\begin{eqnarray}
C(t) &=&\sum_{\vec{x}} \langle 0\mid J_{B}(\vec{x},t) J_{B}^{\dag}(\vec{0},0)
\mid 0\rangle\nonumber\\
&\rightarrow& Cte \exp(-\cal{E} t)
\end{eqnarray}
where $J^{\dag}_{B}$ creates a B--meson from the vacuum.
It is tempting to define $\labar\equiv \cal{E}$. However, the "bare"
binding energy $\cal{E}$ cannot be a physical quantity because it diverges
linearly as $a \rightarrow 0$, $a$ being the lattice spacing \cite{mms}.
The origin of the linear divergence is the mixing of the operator
$\bar{h} D_{4} h$ with the lower dimensional operator $\bar{h} h$. As
pointed out in \cite{mms}, the linear divergence must be subtrated
non--perturbatively in order not to reintroduce renormalon ambiguities.
This divergence can be eliminated from any correlation function by
defining a renormalized action $\bar{h} D^{R}_4 h$ of the form \cite{gmdef}
\be
\bar{h}D^{R}_{4} h = \bar{h} D_{4} h - \left(1-\exp(-a\delta \bar{m})
\right)\, \bar{h}h
\ee
where $\delta \bar{m}$ is a mass counter--term which can be chosen in many
different ways.
Our preferred definition of $\delta \bar{m}$ is \cite{lb1}
\be \label{eq:defdm}
-\delta \bar{m}\equiv \lim_{t\rightarrow \infty}\, \mfrac{1}{a}\,
\ln \left(\mfrac{S(\vec{x},t+a)}{S(\vec{x},t)}\right)
\ee
where $S(\vec{x},t)$ is the heavy--quark propagator in the landau gauge
which can be computed in a numerical simulation of the HQET.
In words, (\ref{eq:defdm}) is the condition that the subtracted heavy--quark
propagator has no exponential fall at large times or equivalently that
$\langle h(p_4=0)\mid \bar{h}D^{R}_{4} h\mid h(p_4=0)\rangle = 0$.
Due to confinement effects, it is not clear that (\ref{eq:defdm}) has a finite
limit as $t\rightarrow \infty$. Our numerical results obtained from three
independent high--statistics simulations at $\beta=6.0,6.2$ and $6.4$
support our assumption on the infrared bahaviour of the heavy--quark
propagator: no confining effects are visible up to distances of about $1.5$
fm, which is the largest distance which we can reach \cite{lb1,lb2}.
An alternative definition of $\delta\bar{m}$ which does not require any
assumption about the large time behaviour of the heavy quark propagator
is given in refs.\cite{lb1,lb2}. However, with this definition $\labar$ is
not a parameter of $O(\Lambda_{QCD})$.
Notice also that the divergent part of $\delta \bar{m}$ must be gauge
invariant. Moreover, the gauge dependence of possible finite parts
is eliminated from physical quantities up to the order of perturbation
theory at which the calculation is performed.
The renormalized binding energy $\labar$ which is of order $O(\Lambda_{QCD})$,
free of renormalon singularities and independent of the
renormalization scale is defined as: $\labar\equiv \cal{E}-\delta \bar{m}$.
Our best estimate of $\labar$ is $\labar=180^{+30}_{-20}$ MeV (see
ref.\cite{lb2} for details).

We can also compute the $\overline{MS}$ running mass of the b--quark,
$\mb$. By comparing the heavy--quark propagator in the continuum with the
one on the lattice, we obtain \cite{lb1}
\be
\mb = \left(M_{B} - \labar\right) \times \overline{C}_{m}(\alpha_{s},
\delta\bar{m})
\ee
where $\overline{C}_{m}$ is the QCD--HQET matching constant
(see ref.\cite{lb1} for
details). Taking $M_{B}=5.278$ GeV, we find $\mb=4.15\pm0.05\pm0.20$ GeV
\cite{lb2},
where the first error is statistical and the second is an estimation of
the error in the truncation at $O(\alpha_{s})$ of the perturbative factor
$\overline{C}_{m}$,
derived from the bubble--resummation
approximation \cite{ms}. This error can be
reduced to about $100$ MeV by computing perturbative
corrections at two--loop order. This calculation requires the
evaluation of the heavy quark propagator in the continuum, already done,
and on the lattice at $O(\alpha_{s}^{2})$. 
The Langevin approach of ref.\cite{rmo} can be used to perform the 
computation on the lattice. 

\section{The calculation of $\lle$.}
\label{kinetic}

Like $\labar$, $\lle$ has no physical meaning since it depends on the
definition that one decides to adopt for the kinetic operator.
On the lattice, matrix elements of this operator contain linear and
quadratic divergences \cite{mms}. The finite kinetic operator, which is
free of power divergences, has the form
\be
\bar{h}\vec{D}_{R}^{2} h = \bar{h}\vec{D}^{2} h - \mfrac{c_1}{a}\,
\bar{h} D^{R}_{4} h - \mfrac{c_2}{a^{2}}\, \bar{h} h
\ee
where $c_1$ and $c_2$ are subtraction constants which we define
by im\-po\-sing the "phy\-si\-cal" non--per\-tur\-ba\-tive 
renormalization prescription \cite{gmdef}
\be
\langle h(\vec{p}=0)\mid \bar{h}\vec{D}^{2}_{R} h\mid h(\vec{p}=0) \rangle
= 0
\ee
in the landau gauge. This is equivalent to the condition
$\rho_{\vec{D}^{2}}(t) = c_1 + c_2\, t$ where
%\be \label{rhorat}
%\rho_{\vec{D}^{2}}(t) \equiv \mfrac{\sum_{\vec{x},\vec{y},t\prime=0}^{t}
%\, S(\vec{x},t\mid\vec{y},t\prime) \vec{D}^{2}_{y}(t\prime)
%   S(\vec{y},t\prime\mid\vec{0},0)}{\sum_{x}\, S(\vec{x},t\mid\vec{0},0)}
%\ee
\be \label{rhorat}
\rho_{\vec{D}^{2}}(t) \equiv \mfrac{\sum_{\vec{x},\vec{y},t'=0}^{t}
\, S(x,y) \vec{D}^{2}_{y}(t')
S(y,0)}{\sum_{\vec{x}}\, S(x, 0)}
\ee
with $x=(\vec{x},t)$ and $y=(\vec{y},t')$.
To determine $c_1$ and $c_2$, we fit $\rho_{\vec{D}^{2}}(t)$ for large $t$
to a straight line. Our numerical results show that the infrared limit
of the ratio in eq.(\ref{rhorat}) does indeed exist,
in spite of the confinement effects in the heavy--quark propagator.
$c_2$ can also be obtained directly by eliminating the sum over $t'$  in
eq.(\ref{rhorat}) and searching for a plateau in $t$ and $t'$. We find
that, after subtracting the contribution of contact terms, both methods give
indistinguishable results (see ref.\cite{l12} for details).
Notice also that, for the same reason as for $\delta\bar{m}$,
$c_2$ must be gauge independent.

The renormalized kinetic energy $\lle$ is then given by:
$\lle=Z_{\vec{D}^{2}}\, \left(\lle^{bare} - c_{2}\right)$, where
$\lle^{bare}=\langle B\mid \bar{h} \vec{D}^{2} h\mid B\rangle /(2 M_{B})$
and $Z_{\vec{D}^{2}}$ is a gauge invariant matching constant known up to
one loop order \cite{mms}. A convenient way to extract $\lle^{bare}$ is
to study the large time behaviour of the ratio of the two--point and the
three--point, with an insertion of the kinetic operator, B--meson
correlation functions. The significant reduction of the statistical errors
with respect to our previous study allows for a much better
identification of the plateau region, which could not be really observed
in ref.\cite{lb1}.
We find $\lle(B_{d})=0.09\pm0.14$ GeV$^{2}$ and
for the physical parameter $\lle(B_{s})-\lle(B_{d})=-0.09\pm 0.04$ GeV$^{2}$
(see ref.\cite{l12} for details).
In order to make a meaningful comparison of our results with other theoretical
determinations, a perturbative calculation of the relation between the
different definitions of $\lle$, and also of $\labar$, is needed at a sufficient
degree of accuracy \cite{ms}. This calculation is missing to date.

\section{The calculation of $\lc$}

The chromo--magnetic operator that enters in the calculation of $\lc$ is
only logarithmically divergent in the ultraviolet cut--off so that there is no
need of a non--perturbative subtraction. Moreover, $\lc$ has a physical
meaning because it corresponds directly to the measurable $B^{*}$--$B$ mass
splitting up to higher power corrections in $1/m_{b}$
\be
M_{B^{*}}^{2}\, -\, M_{B}^{2}\, =\, 4\, \lambda_{2}\, =\, 4\, Z_{\vec{\sigma}
\cdot \vec{G}}\, \lambda_{2}^{bare}
\ee
where $Z_{\vec{\sigma}\cdot\vec{G}}$ is the renormalization constant
necessary to remove the logarithmic divergence present in the bare
operator. The procedure to extract $\lambda_{2}^{bare}$ is the same as for
$\lambda_{1}^{bare}$ (see ref.\cite{l12} for details).
We find $\lc = 0.07 \pm 0.01$ GeV$^{2}$ corresponding to a mass splitting
which is about one half of the experimental value. In our opinion,
there are three possible explanations of the discrepancy, which is
common to all lattice results. The first is the
fact that, at one loop order, $Z_{\vec{\sigma}\cdot\vec{G}}$ is very large
indicating that higher--order terms may modify this factor significantly.
A second source of systematic error is quenching. The final possibility is
that the discrepancy is due to the fact that the mass splitting receives
significant contributions from higher terms in the HQET. Further investigation
is needed on this subject.

\section{The calculation of $B_{B}$ and $f_{B}$}

We have also
computed the parameter $f_{B}^{2} B_{B}$, which enters the theoretical
prediction of the $B^{0}$--$\bar{B}^{0}$ mixing parameters
$x_{d}$ and $x_{s}$. We measure $f_{B}$ and $B_{B}$ from the ratios of
three-- and two--point functions with the $\triangle B=2$
operator fixed at the origin.
Due to the high statistics we have accumulated, we can study
this ratios at large time distances at which the ground state has
been unambiguosly isolated. We find a very good signal for all the
correlation functions and ratios. However, the main uncertainty in our 
results comes from the evaluation of
the renormalization factors relating the effective operators on the lattice
to those in the continuum.
For $B_{B}$, we use the new calculation of the NLO Wilson coefficients
of the $O(\alpha_{s})$ full theory--HQET matching, which takes into account
previously missed contributions \cite{cfg}.
In our opinion, a non--perturbative computation of these factors
is crucial to reduce the rather big systematic errors. This work is
in progress now \cite{gmst}.
Our best estimates are \cite{gm}: $\hat{B}_{B_{d}}=1.21\pm 0.06$
(RGI B parameter),
$\hat{B}_{B_{s}}/\hat{B}_{B_{d}}=1.011\pm0.008$ and
$f^{2}_{B_{s}} B_{B_{s}}/f^{2}_{B_{d}} B_{B_{d}}=1.38\pm 0.07$, where we
use the boosted perturbation theory to evaluate the renormalization
constants.

\end{document}